\documentclass[12pt]{article}
\usepackage{amsbsy,graphicx,latexsym}
\newcommand{\beq}{\begin{equation}}
\newcommand{\eeq}{\end{equation}}
\newcommand{\bea}{\begin{eqnarray*}}
\newcommand{\eea}{\end{eqnarray*}}
\newcommand{\beaq}{\begin{eqnarray}}
\newcommand{\eeaq}{\end{eqnarray}}
\catcode`\@=11 \@addtoreset{equation}{section}

\catcode`\@=11
\def\section{\@startsection{section}{1}{\z@}{3.5ex plus 1ex minus
   .2ex}{2.3ex plus .2ex}{\large\bf}}

\begin{document}
\begin{flushright}
KIAS-P04001 \\ hep-th/0401008
\end{flushright} 

\vspace{10mm}
\centerline{\Large \bf 
Scattering theory of space-time non-commutative}
\centerline{\Large \bf 
abelian gauge field theory}
\vskip 1cm
\centerline{\large Chaiho Rim$^{1}$ and Jae Hyung Yee$^{2}$}
\vskip 1cm 
\centerline{\it $^{1}$ Department of Physics, Chonbuk National University}
\centerline{\it Chonju 561-756, Korea}
\centerline{\it rim@mail.chonbuk.ac.kr}
\vskip .5cm
\centerline{\it $^{2}$ Institute of Physics and Applied Physics, Yonsei University}
\centerline{\it Seoul 120-749, Korea}
\centerline{\it jhyee@phya.yonsei.ac.kr}
\vskip 2cm

\centerline{\bf Abstract}
\vskip 0.5cm
\noindent 
The unitary S-matrix for the space-time non-commutative QED is 
constructed using the $\star$-time ordering 
which is needed in the presence of derivative interactions. 
Based on this S-matrix, perturbation theory is formulated 
and Feynman rule is presented. 
The gauge invariance is explicitly checked 
to the lowest order, 
using the Compton scattering process.
The gauge fixing condition
dependency of the classical solution
of the vacuum is also discussed.

\newpage

\section{Introduction:}
\noindent

Non-commutative field theory (NCFT) \cite{sw,szabo} is the field theory 
on the {\sl non-commutative} (NC) coordinates space,
\beq
\label{nccoord}
[x^\mu, x^\nu] = i \theta^{\mu\, \nu} \,.
\eeq
Space non-commutative theory (SSNC) involves
only the space non-commuting coordinates
( $\theta^{0\nu} =0$),
whereas space-time non-commutative theory (STNC)
contains the non-commuting time 
( $\theta^{0\nu} \ne 0$).
NCFT is constructed based on the Weyl's idea \cite{weyl}:
Instead of using this non-commuting coordinates directly,
one may use the $\star$-product of fields 
over {\it commuting} space-time coordinates.
The $\star$-product encodes all the non-commuting 
nature of the theory and fixes the ordering ambiguity 
of non-commuting coordinates.

We adopt the Moyal product \cite{moyal}
as the $\star$-product representations, 
\beq 
f\star g\, (x)=
e^{\frac{i}{2}{\partial}_x \wedge 
{\partial}_y}f(x)g(y)|_{y=x} \,
\eeq
where $a \wedge b = \theta^{\mu\nu} a_\mu b_\nu$. 
$\theta^{\mu\nu} $ is an antisymmetric c-number 
representing the space-time non-commutativeness.
Using this idea, the commutator in (\ref{nccoord})
becomes the $\star$-commutator,
\bea
[x^\mu \stackrel{\star}, x^\nu] 
\equiv 
x^\mu \star x^\nu - x^\nu \star x^\mu
= i \theta^{\mu\, \nu} \,,
\eea
where the coordinates $x$'s are treated as 
the commuting ones.  
The merit of the Moyal product is that 
the $\star$-product maintains
the ordinary form of the kinetic term of the action,
and allows the conventional perturbation 
where the interaction terms 
become non-local reflecting the non-commuting nature
of the interaction. 

In \cite{ry,ptscalar}
the unitary S-matrix was constructed 
using Lagrangian formalism of the second 
quantized operators in the Heisenberg picture.
Here fixing the time-ordering ambiguity 
has the central role in establishing the unitary S-matrix.
The solution is given as the so-called minimal realization
of the time-ordering step function and $\star$-time ordering. 
The proposal by \cite{bahns} 
to avoid the unitarity problem \cite{nu}
in STNC is in the right direction but the time-ordering 
suggested in \cite{bahns} needs higher derivative correction.
After the higher derivative correction,
the right time-ordering turns out 
to be the $\star$-time ordering as given in \cite{ry,ptscalar}.
Based on this S-matrix, 
the perturbation theory of STNC 
is illustrated using the real scalar theory
and Feynman rule is presented.  

In this paper, we continue 
this project to investigate 
the STNC $U(1)$ gauge field theory.
It is well-known in the {\it commutative} gauge theory
that the proper time-ordering 
is the covariant time-ordering 
due to the derivative interaction.
In this sense, the time-ordering in  STNC gauge theory 
needs to be modified from the one defined 
in the STNC scalar field theory,
whose commutative version contains 
no derivative interaction. 

In addition, the gauge theory possesses the gauge symmetry
and its quantized version should maintain the gauge 
symmetry to get rid of the unphysical states from
the Hilbert space. 
On the other hand,  
it was pointed out that so-called 
the time ordered perturbation theory (TOPT) proposed 
in \cite{topt} does not preserve the gauge symmetry. 
Therefore, for the STNC gauge theory to make sense,
one needs to construct the S-matrix,
not only unitary but also gauge invariant. 

In section \ref{S},  unitary 
S-matrix  of NC abelian gauge theory is presented,
with the proper time-ordering. Due to the presence of the
derivative interaction, the covariant time-ordering 
is necessary in addition to the minimal realization of
the $\star$-time-ordering in scalar theories.
In section \ref{F}, 
Feynman rule is presented in the momentum space.
In section \ref{Q}, STNC quantum electrodynamics is considered
and Feynman rule is presented. 
In section \ref{C}, the gauge invariance is explicitly
checked to the lowest order 
using the Compton scattering amplitude.
This non-trivial check provides how 
the $\star$-time ordering 
cures the defects present in the TOPT in \cite{topt}.
In section \ref{V}, classical vacuum solutions 
is re-analyzed for the STNC pure gauge theory
using the various gauge fixing conditions, 
and how the gauge fixing conditions affect the 
vacuum solution.
Section \ref{L} is the conclusion. 

\vskip 0.5cm

\section{S-matrix of NC abelian gauge theory \label{S}}
\noindent

The NC $U(1)$ action in D-dimensional space-time is given as
\beq 
S = - \frac1{4} \int d^D  x 
F_{\mu \nu}\star F_{\mu\, \nu}
\label{action}
\eeq
where $F_{\mu \nu}$ 
is the field strength,
\bea
F_{\mu \nu}  = 
\partial_\mu A_\nu - \partial_\nu A_\mu 
-ig [A_\mu \stackrel{\star}, A_\nu ] 
\eea
with the gauge coupling constant $g$. 
The action is gauge invariant 
under the gauge transformation,
\bea
ie A_\nu ^\prime(x) = 
U (x)\star \Big(\partial_\mu -ig A_\mu (x)\Big) \star
\bar U(x) \,.
\eea 
This shows that the field strength is 
gauge-covariant rather than gauge-invariant.
For the in-coming or out-going 
photon, however, the field strength is gauge invariant, 
since according to the fundamental ansatz of the 
field theory, the in- or out-photon 
is assumed to subject to the free theory which is 
the commutative field theory. 

As noted in the scalar theory case, the non-commutative 
nature of space and time does not allow 
the unitary transformation of the quantum field 
in the intermediate time into free theory.
Nevertheless, S-matrix can be defined, which relates 
the out-going field to the in-coming field.
Therefore, to construct S-matrix 
we do not need to transform the field strength 
in NCFT into the abelian commutative field strength
as in SSNC gauge theory \cite{sw}. 

With the Lorentz gauge fixing, the action is modified as
\beq
S= - \int d^D x \Bigg\{
\frac14 F_{\mu \nu}\star F^{\mu\, \nu} 
+ \frac{\lambda}2 
(\partial_\mu A^\mu) \star
(\partial_\nu A^\nu) \Bigg\}\,.
\label{g-action}
\eeq
This action can be rewritten in terms of the 
star-operation $\cal F$, 
\beq
S= \int d^D x \Bigg\{ {\cal K}(x) + 
{\cal F}_x \Big({\cal V}(x) \Big) \Bigg\}
\eeq
where $\cal K$ is the usual kinetic Lagrangian density, 
\beq
{\cal K}(x) = 
\frac12 A^\mu \Big(
\partial^2 \eta_{\mu\nu} 
-\partial_\mu \partial_\nu (1 - \lambda)\}
\Big) A^\nu (x) \,.
\eeq
${\cal V} (x) $  is the interaction Lagrangian density 
before the star-operation, 
whose form should be written as the non-local form 
since the split space-time coordinates should be kept:
\bea
&& {\cal V}  (x) \equiv 
{\cal W}_3 (x_1,x_2,x_3)+ {\cal W}_4 (x_1,x_2,x_3,x_4)
\\
&& {\cal W}_3 (x_1, x_2,x_3)
=  g \, \Bigg(
B_{\mu\nu} (x_1)  C^{\mu\nu } (x_2,x_3) 
+ C^{\mu\nu } (x_1, x_2 ) B_{\mu\nu} (x_3)  \Bigg)  
\\&& 
{\cal W}_4 (x_1,x_2,x_3,x_4)
= g^2  C_{\mu\nu} (x_1,x_2) C^{\mu\nu}  (x_3,x_4) \,,
\eea 
where 
\bea
B_{\mu\nu}(x) 
&=& \frac i2 \Big(
\partial_\mu A_\nu (x) -\partial_\nu A_\mu (x) \Big)
\\  
C_{\mu\nu} (x_1,x_2) 
&=&  \frac 12  \Big( A_\mu (x_1)  A_\nu (x_2)  
- A_\nu (x_1) A_\mu  (x_2) \Big) 
=-C_{\nu\mu} (x_1,x_2) \,.
\eea
Here the coordinates $x_i$'s are the split 
coordinates from $x$ and 
the ordering of the split coordinates 
are kept in each term. 
This is needed to have the desired starred action 
after star-operation applied to ${\cal V} (x) $.
The star-operation $\cal F$ is explictly given as 
\beaq
 && \!\!\!\!\!\!
{\cal F}_x \Bigg( {\cal V } (x) \Bigg)  
=  \Bigg\{ 
 e^{ \frac i2 \Big(\partial_{x_1} 
\wedge (\partial_{x_2} + \partial_{x_3})
+ \partial_{x_2} \wedge  \partial_{x_3} \Big)} 
{\cal W}_3 (x_1,x_2,x_3)
\Bigg\}_{x_1=x_2=x_3=x}
\cr 
 && +
\Big\{  e^{\frac i2 \Big(\partial_{x_1} 
\wedge (\partial_{x_2} + \partial_{x_3} + \partial_{x_4})
+ \partial_{x_2} \wedge ( \partial_{x_3}  + \partial_{x_4})
+ \partial_{x_3} \wedge \partial_{x_4} \Big)}
{\cal W}_4 (x_1,x_2,x_3,x_4)
\Bigg\} _{x_1=x_2=x_3=x_4=x}
\!\!\!\!\!\!\!\!\!\!\!\!\!\!\!\!\!
\,.\qquad\qquad\qquad
\eeaq
The Heisenberg equation of motion is given as
\beq
\{ \partial^2 \eta^{\mu\nu} 
-\partial^\mu \partial^\nu (1 - \lambda)\}
A_\nu (x) = \xi^\mu_\star (x)
\equiv \int d^D y\, {\cal F}_y 
\Bigg( \xi^\mu (x;y)\Bigg) 
\label{heisenberg}
\eeq
where 
\bea
\xi(x;y) = \frac{\delta }{\delta A_\mu (x)} 
 {\cal V}(y) \,.
\eea
Here the ordering of the gauge field 
operators should be taken care of. 
The best of this ordering can be done 
in term of the delta-function, 
whose details can be found in \cite{ptscalar}.

The solution of (\ref{heisenberg}) is formally written 
in terms of an integral equation, 
\beaq
A_\mu (x) 
&=&a_\mu (x) + D_{\rm ret}  
\circ \xi_\star ^\nu (x)
\nonumber\\
&=&b_\mu (x) + D_{\rm ad} 
 \circ \xi_{\star \, \mu} (x)\,,
\label{A}
\eeaq
where
$a_\mu(x)$ and $ b_\mu(x)$ are 
in- and out-fields: 
\bea
a_\mu(x)  \equiv A_{\mu\, \rm in}(x) \,,\qquad
b_\mu(x)  \equiv A_{\mu\, \rm out} (x) \,,
\eea
and $\circ$ denotes the convolution, 
\bea
G \circ \xi_\star ^\nu\, (x) 
&=& \int d^D y \,\, 
G(x-y) \,\,\xi_\star ^\nu (y) \,.
\eea
The retarded and advanced Green's functions are given 
in terms of the free commutator function of 
massless scalar field
$D(x)$,
\bea
&&D_{\rm ret}(x)  
= -\theta(x^0) \,D (x) \,,\quad
D_{\rm ad} (x)  
= \theta(-x^0) \, D (x)  \,,\quad
\\
&& D(x) 
= \int \frac{d^D k}{(2\pi)^D} \,\, e^{-ikx} \,
2\pi  \delta(k^2 )\, \epsilon (k^0) \,.
\eea
Here we take the gauge parameter $\lambda =1$ (Feynman gauge)
to avoid the unnecessary complication.

It is worth to mention the effect of the 
derivative interaction. 
The commutative abelian gauge theory 
contains the derivative interaction terms
and therefore, $\xi^\mu (x)$ contains the derivatives of 
delta functions in addition to the derivatives due to the 
direct $\star$-operation. 
This derivative of the delta function will 
result in the derivative of the retarded (or advanced) 
Green's function after the convolution integral :
\beaq
G \circ \bigg(
A(x) \frac{\partial \delta(x)}{\partial x_\mu }\bigg)
&=& \int d^D y \,
A(y) \,\frac{\partial G(x-y)} {\partial y_\mu }\,.
\label{derivative}
\eeaq
Then it is obvious that 
the derivative of the retarded (or advanced) 
Green's function will bring in the contact term 
through the derivative of the step-function. 
This result will have the important effect 
on the S-matrix later on, especially in time-ordering. 

The in-coming gauge field is assumed to satisfy 
the free commutation relation,
\bea
[\, A_{\mu \, \rm free} (x),\, 
A_{\nu\, \rm free} (y)\,] = - i \eta_{\mu\nu}\, D(x-y)\,.
\eea
The quantized free gauge field is expressed as
\bea
a_\mu (x)
\equiv 
A_{\mu \rm free} (x) 
= \int \frac{d ^{D-1} \bf p}{(2\pi)^{D-1}}
\frac 1{2 |{\bf p}|} \sum_{\lambda} \Bigg(
a({\bf p}, \lambda) \epsilon_\mu ({\bf p}, \lambda)
e^{-i p \cdot x} +
a({\bf p}, \lambda)^\dagger \,
\bar \epsilon_\mu ({\bf p}, \lambda)
e^{i p \cdot x} \Bigg)
\eea
where  $\bf p$ represents the $D-1$ dimensional momentum.
The creation and annihilation operators satisfy the 
commutation relation, 
\bea 
[\,a( {\bf p},  \lambda),\,  a ^\dagger ({\bf q} , \lambda')\,]
= - (2\pi)^{D-1} \, 2 |{\bf p}|\,  
\delta ({\bf p} -{\bf q}) \,
\eta^{\lambda\lambda'}\,,
\eea
and the polarization vector satisfies the relation
\bea
\sum_{\lambda} \epsilon_\mu ({\bf k}, \lambda)\,
\bar \epsilon_\nu ({\bf k}, \lambda) 
= \eta_{\mu\nu} \,,\qquad
\epsilon_\mu ({\bf k}, \lambda)\,
\bar \epsilon^\mu ({\bf k}, \tau ) 
= \eta^{\lambda\tau}  \,.
\eea

The integral equation (\ref{A}) is assumed to be solved 
iteratively in terms of the in-field $a_\mu (x)$,
and, therefore, $\xi^\nu (x) $ can be written in terms of 
the in-fields only. 
As a result, the out-field 
is written in terms of the in-field:
\beq
b^\mu(x)  = a^\mu(x) -  D  
\circ \xi_{\star}^\mu (x) \,.
\eeq
If the out-field
is written in terms of an S-matrix 
\beq
b^\mu (x) 
=  S^{-1} \,  a^\mu (x) \,\,S \,,
\label{inoutSmatrix}
\eeq
then, out-field also respects 
the free commutator relation. 

One may repeat the same procedure given 
in the scalar case to construct the S-matrix.  
However, two additional points
are to be considered carefully in the gauge theory;
derivative interaction terms and the gauge symmetry. 
All of them are not new in this 
starred interaction 
but are inherent to the gauge theory.
The solution to this two ingredients 
are already considered 
in the commutative gauge theory 
and therefore, we need to check 
if the solution works correctly.

The S-matrix is written as 
\beaq    
S &=&  1 +i \int_{-\infty}^\infty d^D x \,{\cal F}_x \,
\Big( {\cal V} (x) \Big)
+ i^2 \int_{-\infty}^\infty \!\int_{-\infty}^\infty 
d^D x_1 d^D x_2 \,
{\cal F}_{12}\, \Big(\theta^*_{12} {\cal V} (x_1) 
{\cal V}(x_2) \Big)
\cdots \nonumber \\ 
&& \quad + 
i^n \int_{-\infty}^\infty \!
\cdots \int_{-\infty}^\infty d^D x_1 \, \cdots \,d^D x_n 
{\cal F}_{1 2\cdots n}\,
\Big( \theta^*_{12 \cdots n} 
{\cal V} (x_1) \cdots {\cal  V}  (x_n) 
\Big) + \cdots   \,,                                
\label{s-matrix}
\eeaq
where $\cal F$ is the star-operation and 
$\theta^*_{12\cdots n}$ is the covariant time-ordering,
whose detailed forms will be given promptly.
The time-integration has to be done  outside of the 
star-operation. We put the space integration 
as well as the time integration 
outside the star-operation 
for notational convenience. 

The covariant time-ordering 
$\theta^*_{12\cdots n}$
is the ordinary time-ordering 
when there is no derivative operators,
which is defined 
in terms of the step function,  
\bea
\theta_{12}^* \,\phi(x_1) \phi(x_2) 
=\theta (t_1 -t_2)\, \phi(x_1) \phi(x_2)\,. 
\eea
The time-ordering of derivative operators, however,
need be done in the form of covariant time ordering:
The covariant time ordering is defined  
before taking derivatives:
\beq
\theta_{12}^* \Big(f(\partial_1) \phi(x_1)\Big) 
\Big( g(\partial_2) \phi(x_2) \Big)
=f(\partial_1 ) g(\partial_2)
\Big( \theta (t_1 -t_2)\phi(x_1) \phi(x_2) \Big) \,. 
\eeq
The composite covariant time-ordering 
is similarly given as
\bea
\theta_{12\cdots n}^* = \theta^*_{12} \,
\theta^*_{23} \cdots \theta^*_{n-1\, n}\,.
\eea
The covariant time-ordering is the source of 
Schwinger's contact term:
\bea
\theta_{12}^* \Big( \partial_\mu j^\mu (x_1)
j^\nu (x_2) \Big)
= \delta (x_1^0-x_2^0) \Big( j^0 (x_1)j^\nu (x_2)\Big) 
+ \theta_{12}\Big( \partial_\mu j^\mu (x_1)
j^\nu (x_2) \Big) \,.
\eea

The composite $\star$-operation 
is defined as 
\bea
{\cal F}_{1 2\cdots n}
\equiv {\cal F}_1  {\cal F}_2 \cdots {\cal F}_n \,,
\eea
which is commutative,
\bea 
{\cal F}_{1 2}
\equiv {\cal F}_1  {\cal F}_2 
= {\cal F}_2  {\cal F}_1 ={\cal F}_{21} \,,
\eea
so that 
\bea  
{\cal F}_{1 2\cdots n}
\Bigg( {\cal V} (x_1) {\cal V} (x_2) \cdots {\cal V} (x_n) \Bigg) 
=  {\cal L}_I (x_1) {\cal L}_I(x_2) \cdots {\cal L}_I (x_n) \,.
\eea

The star-operation $\cal F$ needs special care
in the presence of the time-ordering step function. 
For example, suppose one considers 
the star-operation on two vertices 
with a time step function,
\bea
{\cal F}_{x y}
\Bigg( \theta^*(x^0-y^0) {\cal Z}(x) 
{\cal Z}(y)\Bigg)\,,
\eea
and the vertices are given 
in terms of quantum fields with no derivatives:  
\bea
{\cal Z}(x)= a_{\mu_1}(x)a_{\mu_2}(x) \cdots a_{\mu_n} (x) \,.
\eea
The star-operation can be done as in the scalar theory,
\bea
&& \!\!\!\!\!\!\!\!\!
{\cal F}_{x y}
\Bigg( \theta^*(x^0-y^0) {\cal Z}(x) {\cal Z}(y)\Bigg)
\\&& = {\cal F}_x \,{\cal F}_y
\Bigg( \theta(x^0 -y^0) \,
\Big( a_{\mu_1}(x_1) a_{\mu_2} (x_2) 
\cdots a_{\mu_n} (x_n) \Big)
\Big( a_{\nu_1}(y_1)a_{\nu_2}(y_2) 
\cdots a_{\nu_n} (y_n) \Big)   
\Bigg|_{x_i = x\,, y_i=y} \Bigg)\,,
\eea
where the minimal realization
is used to avoid the ambiguity of the step-function.
The minimal realization is to put
the step function $\theta(x^0 -y^0)$ 
as $ \theta(x_i^0 -y_j^0) $ {\sl only once} 
if the operators $a_{\mu_i}(x_i)$ and $a_{\nu_j}(x_j)$ 
are contracted. 
Even in the presence of many spectral functions 
the time-ordering step function 
should be used {\sl only once}
between two vertices: 
\bea
\theta(x^0 -y^0) \prod_{i, j}  \Delta(x_i -y_j) 
\to \theta(x_a^0 -y_b^0) \prod_{i, j}  \Delta(x_i -y_j) 
\eea
where $a$ ($b$) is just one of indices among $i$'s ($j$'s). 
In addition, for the connected $n$-vertices, there are only
$n$-number of step functions. 
This minimal realization originates from the 
consistency condition of S-matrix 
(\ref{inoutSmatrix}) and Heisenberg equation 
of motion (\ref{heisenberg}). The details are given 
in \cite{ptscalar}. 
When the vertices contain derivatives, 
one needs the mininal realization
with the covariant time-ordering. 

The proof that the S-matrix in (\ref{s-matrix})
relates the in- and out-field  (\ref{inoutSmatrix})
and is unitary  $S^\dagger S =1$ 
goes the same as in the scalar theory,
if one takes into consideration  
the effect of derivative interaction:
The derivative interaction 
necessitates the covariant time-ordering 
as noted in ({\ref{derivative}).

Introducing a notation of the 
time-ordering $\star$-product as,
\beq
T_{\star} \{{\cal V}(t_1) {\cal V}(t_2)\} = 
{\cal F}_{12} \Big( \theta_{12}^* \, {\cal V}(t_1) {\cal V} (t_2)
+ \theta_{21}^* \, {\cal V}(t_2) {\cal V} (t_1)\Big)\,.
\eeq 
we can put the S-matrix in a compact form as 
\beaq
S &=& \sum_{n=0}^\infty \frac{i^n}{n!}
\int_{-\infty}^\infty d^Dx_1 
\cdots \int_{-\infty}^\infty d^Dx_n\,
T_{\star}\, \Big\{{\cal V} (\phi_{\rm in}(x_1))
\cdots  {\cal V} (\phi_{\rm in}(x_n) \Big\}
\nonumber\\
&\equiv& T_{\star} \,
\exp \Bigg( i\int_{-\infty}^\infty dx \,
{\cal V}(\phi_{\rm in}(x) \Bigg)\,.
\eeaq

Finally, the gauge symmetry of the system (\ref{action})
requires the scattering amplitude to be gauge invariant. 
However, the modified action with the addition of the gauge fixing term
(\ref{g-action}) allows the non-physical degree of freedom
and one needs to check the non-physical degree of freedom
does not affect the scattering amplitude. 
The is achieved by defining the physical states,
$ |{\rm phy} \rangle $ in a weak form,
\bea
\langle {\rm phy} 
|\,\partial_\mu a^\mu |{\rm phy}\rangle =0 
\eea
or by introducing Faddeev-Popov ghost fields into the system.
The gauge invariance of the scattering amplitude 
will be considered in section \ref{C}.

\vskip 0.5cm
\section{Feynman Rule \label{F}}
\noindent

NC abelian gauge theory contains 
the derivative interaction and 
the 3-point and 4-point vertex function is given 
at the lowest order as following :
\beaq
\langle a_\mu(p_1) a_\nu(p_2) a_\rho(p_3) \rangle_c
&\equiv&  -i g (2\pi)^d \delta^d (p_1 + p_2 +p_3 ) 
\,\, v_{\mu\nu\rho}(p_1,p_2,p_3) 
\cr
\langle a_\mu(p_1) a_\nu(p_2) 
a_\rho(p_3)a_\sigma(p_4)  \rangle_c
&\equiv & - ig^2 \,
(2\pi)^d \delta^d (p_1 + p_2 +p_3 +p_4) 
\,\, v_{\mu\nu\rho\sigma} (p_1,p_2,p_3,p_4) 
\qquad\quad
\eeaq
where $\langle \cdots \rangle_c$ refers to the
1-particle irreducible vertex and 
\beaq
 v_{\mu\nu\rho}(p_1,p_2,p_3) 
&=&
-2i \sin \Big( \frac{ p_2 \wedge p_3}2 \Big) 
\Big( (p_1)_\nu \delta_{\mu \rho} -  \nu \leftrightarrow \rho \Big) 
\cr
&&
\qquad
+ \Big((p_1,\mu) \leftrightarrow (p_2, \nu)\Big)
+ \Big((p_1,\mu) \leftrightarrow (p_3, \rho)\Big)
\cr
 v_{\mu\nu\rho\sigma} (p_1,p_2,p_3,p_4) 
&=&
-\sin\Big( \frac {p_1 \wedge p_2}2 \Big)
\sin \Big(\frac {p_3 \wedge p_4}2 \Big) 
\Big( \delta_{\mu\rho} \delta_{\nu \sigma} -  \sigma \leftrightarrow \rho \Big) 
\cr
&&\qquad\qquad\qquad
+ \Big((p_2,\nu) \leftrightarrow (p_3, \rho)\Big)
+ \Big((p_2,\nu) \leftrightarrow (p_4, \sigma)\Big)\,.
\qquad
\eeaq
Here all the momenta are assumed to be in-coming. 

The free 2-point function is given 
in terms of the spectral function
$D^+_{\mu\nu}(x_1,x_2)$ whose Fourier transformed form is
\beq
D^+_{\mu\nu}(x_1, x_2) 
=  \langle a_\mu(x_1) a_\nu(x_2) \rangle 
=  -\eta_{\mu\nu}  D_+ (x_1-x_2)
\eeq
where $D_+(x)$ is the  massless scalar spectral function,
\bea
D_+ (x) 
= \int \frac{d^D p} {(2\pi)^D}
e^{-ip\cdot x} \, 
2\pi \delta(p^2 ) \theta(p^0) \,.
\eea
Therefore, the 2-point function in momentum space is given as 
\beq
\tilde D^+_{\mu\nu}(p) =  -\eta_{\mu\nu}  
2\pi \delta(p^2 ) \theta(p^0) \,.
\eeq

Time-ordered spectral function is defined as 
\beq
D^R_{\mu\nu}(x_1,x_2)
= \theta(x_1^0- x_2^0) D^+_{\mu\nu}(x_1,x_2)
=  \eta_{\mu\nu}   D_R (x_1-x_2)
\eeq
where $D_R (x)$ 
is the massless scalar time-ordered spectral function 
 \bea
D_R (x)
= \int \frac{d^D p} {(2\pi)^D} \;
e^{-ip\cdot x} \, 
\frac {i}{ 2 |{\bf p}|\,
(p^0 -|{\bf p}| + i \epsilon) } \,.
\eea
In momentum space the time-ordered spectral function
of photon is given as
\bea
\tilde D^R_{\mu\nu}(p) =  \eta_{\mu\nu} \, 
\frac {i}{ 2 |{\bf p}|\,(p^0 -|{\bf p}| + i \epsilon )} \,.
\eea
Note that the Feynman propagator is given 
in terms of the time-ordered spectral function as
\beq
i \Delta_{F\, \mu\nu} (x) = -\eta_{\mu\nu}
\Big(D_R (x) + D_R(-x)\Big)\,.
\eeq

Our Feynman rule for this theory 
is summarized as follows.\\
(1) Each vertex has either 3 legs or 4 legs.
3-leg vertex is assigned as $ v_{\mu\nu\rho}(p_1,p_2,p_3)$
and 4-leg vertex as 
$ v_{\mu\nu\rho\sigma}(p_1,p_2,p_3,p_4)$,
where $p_i$ is the in-coming momentum of each leg.
The total momentum at each vertex vanishes.
\bea
\begin{picture}(100,40)(0,20)
\multiput(30,14)(0,12){3}{\oval(1,6)[l]}
\multiput(32,20)(0,12){3}{\oval(1,6)[r]}
\multiput(0,9.5) (12,0){6}{\oval(6,1)[b]} 
\multiput(6,11) (12,0){6}{\oval(6,1)[t]} 
\put(-15,0){$p_1$}
\put(-15,15){$\mu$}
\put(5,7){$>$}
\put(70,0){$p_2$}
\put(70,15){$\nu$}
\put(45,7){$<$}
\put(20,50){$\rho$}
\put(35,50){$p_3$}
\put(27,30){$\vee$}
\end{picture}
&=&  -ig  v_{\mu\nu\rho}(p_1,p_2,p_3)
\\ 
\begin{picture}(100,80)(0,40)
\multiput(29.5,4)(0,12){6}{\oval(1,6)[l]}
\multiput(31,10)(0,12){6}{\oval(1,6)[r]}
\multiput(0,40) (12,0){6}{\oval(6,1)[b]} 
\multiput(6,42) (12,0){6}{\oval(6,1)[t]} 
\put(-15,30){$p_1$}
\put(-15,45){$\mu$}
\put(5,37){$>$}
\put(70,30){$p_2$}
\put(70,45){$\nu$}
\put(45,37){$<$}
\put(20,80){$\rho$}
\put(35,80){$p_3$}
\put(27,58){$\vee$}
\put(20,0){$\sigma$}
\put(35,0){$p_4$}
\put(27,15){$\wedge$}
\end{picture}
&=& - ig^2  v_{\mu\nu\rho\sigma}(p_1,p_2,p_3,p_4)
\\ \\
\eea
(2) The legs at each vertex 
are either external legs or are 
connected to other vertex, making internal lines.\\
(3) Each vertex is numbered so that the internal lines are 
assigned with arrows. 
The arrows point from high-numbered vertex 
to low-numbered one.\\
(4) Among the arrows, 
one arrow between adjacent vertices 
can be assigned as triangled arrow,
with the condition that 
the total number of the triangled arrows should be 
$n-1$ for the $n $ connected vertices  ($n>1$).
This restriction is due to the minimal realization. \\
(5) The diagrams with the same topology with arrows 
are identified and the numbering of vertices 
is ignored. As a result the number of diagrams 
are reduced from the $n!$ diagrams.
In this sense, the numbering of the vertices
are considered as the intermediate step
to obtain the topologically 
distinct diagrams with arrows.\\
(6) The distinctive Feynman diagrams 
are multiplied by the symmetric factors. \\
(7) The momentum flows along the arrows.
The arrowed internal line with momentum $k$ 
is assigned as 
$ \tilde D^+_{\mu\nu} (k)$ and
the triangled arrowed internal line as 
$\tilde D^R_{\mu\nu} (k)$. 
\bea
\begin{picture}(80,20)(0,5)
\multiput(0,9.5) (12,0){6}{\oval(6,1)[b]} 
\multiput(6,11) (12,0){6}{\oval(6,1)[t]} 
\put(-3,10){\circle*{4}}
\put(70,10){\circle*{4}}
\put(35,6){$>$}
\put(-15,15){$\mu$}
\put(20,0){$p$}
\put(70,15){$\nu$}
\end{picture}
&=& \tilde D^+_{\mu\nu} (p)
\\
\begin{picture}(80,20)(0,5)
\multiput(0,9.5) (12,0){6}{\oval(6,1)[b]} 
\multiput(6,11) (12,0){6}{\oval(6,1)[t]} 
\put(-3,10){\circle*{4}}
\put(70,10){\circle*{4}}
\put(35,6){$\triangleright$}
\put(-15,15){$\mu$}
\put(20,0){$p$}
\put(70,15){$\nu$}
\end{picture}
&=& \tilde D^R_{\mu\nu} (p)\,.
\eea

\vskip 0.5cm
\section{STNC QED \label{Q}}
\noindent

When the Dirac fermions are  added to the 
abelian gauge theory,
there appears an additional action,
\beq
\Delta S = \int d^D x 
\Big\{ \bar \psi \, \star 
\,(i \partial\!\!\!/  + m - e A\!\!\!/ )
\star \psi  \Big\}\,.
\eeq 
$A\!\!\!\slash$ is the abbreviation of 
$A^\mu\gamma_\mu$. The gauge coupling
$g$ is identified as the charge of the electron,
$g=-e$. 

The in-coming Dirac fermionic fields
$\psi$ and $\bar \psi$ 
satisfy the free anti-commutation relation, 
\beaq
i S_{\alpha \beta} 
&=& \{\psi_{\alpha}\,, \bar\psi_{\beta}\,\} 
=\int \frac{d^D k}{(2\pi)^{D}}
e^{-i k\cdot x} ( k\!\!\!\slash  + m )_{\alpha \beta} 
2 \pi \delta(k^2 -m^2) \epsilon(k^0)
\cr
&=& (i \partial \!\!\!\slash + m)_{\alpha \beta}\, 
\Big( i \Delta (x) \Big) 
\eeaq
where $i\Delta (x)$ 
is the massive scalar commutator function.
The free fermionic field is written as 
\beaq
\psi(x) &=& 
\int \frac{d^{D-1}  p}{(2 \pi)^{D-1}} 
\, \frac m{\omega_p} \sum_i 
\Big( 
b_i (p) u^i (p) e^{-ip\cdot x}
+ d_i^\dagger (p) v^i (p) e^{ip\cdot x}
\Big)
\cr
\bar \psi (x) &=& 
\int \frac{d^{D-1}  p}{(2 \pi)^{D-1}} 
\, \frac m{\omega_p} \sum_i 
\Big( 
b_i^\dagger (p) \bar u^i (p) e^{ip\cdot x}
+ d_i (p) \bar v^i (p) e^{-ip\cdot x}
\Big)
\eeaq
where $b$ and $d$ are 
($b^\dagger$ and $d^\dagger$)
are annihilation (creation ) operators,
\beaq
\{b_\alpha (p), b_\beta^\dagger(q) \} 
&=& (2\pi)^{D-1} \delta({\bf p}
 - {\bf q}) \,\delta_{\alpha\beta}
\cr
\{d_\alpha (p), d_\beta^\dagger(q) \} 
&=& (2\pi)^{D-1} \delta({\bf p}
 - {\bf q})\, \delta_{\alpha\beta}\,,
\eeaq 
and $u$ and $v$ are spinors
and $\bf p$ denotes the spatial momentum of $p$.

The retarded and advanced Green's function 
are written as
\beq
\Big(S_{\rm ret}\Big)_{\alpha\beta} (x) = -\theta(x^0) S_{\alpha \beta} (x)
\qquad
\Big(S_{\rm adv}\Big)_{\alpha\beta} (x) = \theta(-x^0) S_{\alpha \beta} (x)
\eeq
since 
\bea
(i \partial \!\!\!\slash -m) S_{\rm ret} (x) =
(i \partial \!\!\!\slash -m) S_{\rm adv} (x) = 
- \delta^D (x)\,.
\eea
As in the scalar and gauge theories,
we solve the equation of motion in terms of the 
retarded Green's function with the in-coming wave.
The S-matrix which relates the in- and out- fields
including fermion field are given as
(\ref{s-matrix})  if one 
includes the fermionic part in the interaction 
Lagrangian density $\Delta \cal V$;
\beq
\Delta { \cal V}= e\, 
\bar \psi  \,  A\!\!\!\slash  \, \psi  
\eeq

To evaluate the S-matrix element in momentum space 
we need 2-point functions for the positive 
(particle contribution) 
and negative frequency part (anti-particle contribution), 
and their time-ordered ones. 
The 2-point functions are given as 
\beaq
S^+ _{\alpha \beta} (x) 
=  \langle 0| \psi_\alpha (x)\, \bar \psi_\beta |0 \rangle 
&=& \int \frac{d^d k}{(2 \pi)^D} e^{-ik\cdot x} 
( k\!\!\!\slash + m)_{\alpha \beta} \, 
2 \pi \delta(k^2 -m^2) \theta(k^0) 
\cr
S^- _{\alpha \beta} (x) 
=   \langle 0| \bar \psi_\beta \, \psi_\alpha (x) |0 \rangle 
&=& \int \frac{d^D k}{(2 \pi)^D} e^{ik\cdot x} 
( k\!\!\!\slash - m)_{\alpha \beta} \, 
2 \pi \delta(k^2 -m^2) \theta(k^0) 
\cr
&=&   \int \frac{d^D k}{(2 \pi)^D} e^{-ik\cdot x} 
(- k\!\!\!\slash - m)_{\alpha \beta} \, 
2 \pi \delta(k^2 -m^2) \theta(-k^0) \,,
\eeaq 
and their time-ordered ones are given as 
\beaq
S^{+R}_{\alpha\beta} (x) 
&=& \theta(x^0) S^+_{\alpha\beta} (x)
= \int \frac{d^D k}{(2\pi)^{D}}
\, e^{-i k\cdot x} \,
\frac{ (\hat k\!\!\!\slash^+  + m )_{\alpha \beta} }
{2 \omega_k} 
\frac{i}{k^0-\omega_k +i \epsilon}
\cr
S^{-A}_{\alpha\beta} (x) 
&=& \theta(-x^0) S^+_{\alpha\beta} (x)
= \int \frac{d^D k}{(2\pi)^{D}}
\, e^{-i k\cdot x} \,
\frac{ (\hat k\!\!\!\slash^-  + m )_{\alpha \beta} }
{2 \omega_k} 
\frac{i}{k^0+\omega_k -i \epsilon}
\eeaq
where 
$\omega_k =\sqrt{{\bf k}^2 +m^2 }$
and 
$\hat k^{\pm} = (\pm \omega_k\,, {\bf k})$.
The Feynman propagator is given in terms of the time-ordered
positive and negative spectral functions, 
\beaq
i S^F_{\alpha \beta}
&=& \theta(x^0) S^+_{\alpha \beta}(x)
- \theta(-x^0) S_{\alpha \beta}^- (x) 
= S^{+R}_{\alpha\beta} (x) -S^{-A}_{\alpha\beta} (x) 
\cr
&=& (i \partial_\mu \gamma^\mu  + m)\, 
( i \Delta_F (x)) 
\eeaq
where $i \Delta_F (x)$ is the massive scalar 
Feynman propagator,
\beq
i \Delta_F (x) 
= \Delta_R (x) + \Delta_R(-x)\,.
\eeq

Feynman rule for the fermionic theory
is similarly given as in the bosonic theory. 
There, however, appear a few differences:\\
(1)The extra $-$ sign should be multiplied each time
two fermionic operators interchange each other
due to the fermionic statistics.\\
(2)There are two types of fermionic internal lines,
a solid  arrowed line for particle propagator
and a dashed one for anti-particle one.
The momentum is assumed to flow along the arrow:
\bea
\begin{picture}(100,20)(-25,10)
\put(0,10) {\line(1,0){60}} 
\put(25,7){\footnotesize{$<$}}
\put(35,13){$k$}
\put(-5,15){$\alpha$}
\put(65,15){$\beta$}
\put(0,10) {\circle*{2}}
\put(60,10) {\circle*{2}}
\end{picture}
&=& \,\tilde S^+_{\alpha\beta}(k)
=( k\!\!\!\slash + m)_{\alpha \beta} \, 
2 \pi \delta(k^2 -m^2) \theta(k^0) 
\\
\begin{picture}(100,20)(-25,10)
\put(0,10) {\line(1,0){60}} 
\put(28,7){$\triangleleft$}
\put(35,13){$k$}
\put(-5,15){$\alpha$}
\put(65,15){$\beta$}
\put(0,10) {\circle*{2}}
\put(60,10) {\circle*{2}}
\end{picture}
&=& \,\tilde S^{+R}_{\alpha\beta}(k)
=
\frac{ (\hat k\!\!\!\slash^+  + m )_{\alpha \beta} }
{2 \omega_k} 
\frac{i}{k^0-\omega_k +i \epsilon}
\\
\begin{picture}(100,20)(-25,10)
\multiput(0,10)(10,0){6} {\line(1,0){8}} 
\put(25,7){\footnotesize{$<$}}
\put(35,13){$k$}
\put(-5,15){$\alpha$}
\put(65,15){$\beta$}
\put(0,10) {\circle*{2}}
\put(60,10) {\circle*{2}}
\end{picture}
&=& \,\tilde S^-_{\alpha\beta}(k)
= (- k\!\!\!\slash - m)_{\alpha \beta} \, 
2 \pi \delta(k^2 -m^2) \theta(-k^0) \,,
\\
\begin{picture}(100,20)(-25,10)
\multiput(0,10)(10,0){6} {\line(1,0){8}} 
\put(28,7){$ \triangleleft$}
\put(35,13){$k$}
\put(-5,15){$\alpha$}
\put(65,15){$\beta$}
\put(0,10) {\circle*{2}}
\put(60,10) {\circle*{2}}
\end{picture}
&=& \,\tilde S^{+R}_{\alpha\beta}(k)
= \frac{ (\hat k\!\!\!\slash^-  + m )_{\alpha \beta} }
{2 \omega_k} 
\frac{i}{k^0+\omega_k -i \epsilon}
\eea\\
(3) The direction of arrows along a 
fermionic line should not be reversed
in the Feynman graphs.
If the arrow points from a high-numbered 
vertex to a low numbered vertex,
the fermionic internal line is assigned as
a solid line. Otherwise, the fermionic line
is dashed.\\
(4) The vertex with two fermionic legs 
is assigned as 
\bea
\langle  \bar \psi_\alpha(p) \psi_\beta(q)
A_\mu(k) \rangle_c
&=& -ie\,\,
(2\pi)^D \delta (p +k-q) \, 
v_{\alpha\beta \,;\mu} (p,q;\, k) 
\eea
where 
\bea
-ie\, v _{\alpha\beta \,;\mu} (p,q;\, k) 
&=& -ie\,\, e^{\frac i2 p\wedge q } 
\, \gamma^\mu_{ \alpha \beta }
= \begin{picture}(100,35)(-25,10)
\multiput(30,14)(0,12){3}{\oval(1,6)[l]}
\multiput(31,20)(0,12){3}{\oval(1,6)[r]}
\put(-10,10) {\line(1,0){90}} 
\put(10,7.5){\footnotesize{$<$}}
\put(50,7.5){\footnotesize{$<$}}
\put(26,26){$\vee$}
\put(-12,0){$p$}
\put(-12,15){$\alpha$}
\put(70,0){$q$}
\put(70,15){$\beta$}
\put(23,50){$\mu\; k$}
\end{picture}\cr
\eea
and its total momentum vanishes $q+k+(-p)=0$. 
There are two more types of the vertex
depending on the fermion legs, particle or antiparticle. 
However, the vertex function is independent of the particle 
species.
\bea
\begin{picture}(100,50)(-15,10)
\multiput(30,14)(0,12){3}{\oval(1,6)[l]}
\multiput(31,20)(0,12){3}{\oval(1,6)[r]}
\multiput(-10,10)(10,0){9} {\line(1,0){7}} 
\put(10,7.5){\footnotesize{$<$}}
\put(50,7.5){\footnotesize{$<$}}
\put(26,26){$\vee$}
\put(-12,0){$p$}
\put(-12,15){$\alpha$}
\put(70,0){$q$}
\put(70,15){$\beta$}
\put(23,50){$\mu\; k$}
\end{picture}
= -ie\, v _{\alpha\beta \,;\mu} (p,q;\, k) 
\cr \cr
\begin{picture}(100,50)(-15,10)
\multiput(30,14)(0,12){3}{\oval(1,6)[l]}
\multiput(31,20)(0,12){3}{\oval(1,6)[r]}
\put(-10,10) {\line(1,0){43}} 
\multiput(36,10)(10,0){4} {\line(1,0){7}} 
\put(10,7.5){\footnotesize{$<$}}
\put(50,7.5){\footnotesize{$<$}}
\put(26,26){$\vee$}
\put(-12,0){$p$}
\put(-12,15){$\alpha$}
\put(70,0){$q$}
\put(70,15){$\beta$}
\put(23,50){$\mu\; k$}
\end{picture}
= -ie\, v _{\alpha\beta \,;\mu} (p,q;\, k) 
\eea

\vskip 1cm
\section{Compton Scattering and Gauge Invariance 
\label{C}}

The S-matrix should be gauge invariant. 
In this section, we investigate 
on the Compton scattering 
to the lowest order to prove that the scattering
amplitude is gauge invariant.
The Compton scattering of an electron 
is given as the following diagrams 
to the lowest order:
\beq
M_{\mu\nu\,;\alpha\beta}(k_1,k_2\,;p,q)
= - e^2 \Big( M^{(1)}_{\mu\nu\,;\alpha\beta}(k_1,k_2\,;p,q)
-M^{(2)}_{\mu\nu\,;\alpha\beta}(k_1,k_2\,;p,q)
\Big)
\eeq
where 
\bea
- e^2 M^{(1)}_{\mu\nu\,;\alpha\beta}(k_1,k_2\,;p,q)
&=& 
\begin{picture}(110,60)(-20,10)
\multiput(15,13)(0,12){3}{\oval(1,6)[l]}
\multiput(17,19)(0,12){3}{\oval(1,6)[r]}
\multiput(45,13)(0,12){3}{\oval(1,6)[l]}
\multiput(47,19)(0,12){3}{\oval(1,6)[r]}
\put(-10,10) {\line(1,0){90}} 
\put(28,7){$\triangleleft$}
\put(-15,0){$p$}
\put(-15,15){$\alpha$}
\put(70,0){$q$}
\put(70,15){$\beta$}
\put(8,50){$\mu\; k_1$}
\put(38,50){$\nu\; k_2$}
\put(2,7){$<$}\put(55,7){$<$} \end{picture}
+\qquad
\begin{picture}(100,60)(0,10)
\multiput(40,15)(-10,10){4}{\oval(10,10)[rt]}
\multiput(50,15)(-10,10){4}{\oval(10,10)[lb]}
\multiput(20,15)(10,10){1}{\oval(10,10)[lt]}
\multiput(10,15)(10,10){2}{\oval(10,10)[rb]}
\multiput(40,45)(10,10){1}{\oval(10,10)[rb]}
\multiput(40,35)(10,10){2}{\oval(10,10)[lt]}
\put(-10,10) {\line(1,0){90}} 
\put(28,7){$\triangleleft$}
\put(-15,0){$p$}
\put(-15,15){$\alpha$}
\put(70,0){$q$}
\put(70,15){$\beta$}
\put(5,53){$\mu\; k_1$}
\put(38,53){$\nu\; k_2$}
\put(2,7){$<$}\put(55,7){$<$}
\end{picture}
\\&& \quad
-\qquad
 \begin{picture}(110,60)(-20,10)
\multiput(15,13)(0,12){3}{\oval(1,6)[l]}
\multiput(17,19)(0,12){3}{\oval(1,6)[r]}
\multiput(45,13)(0,12){3}{\oval(1,6)[l]}
\multiput(47,19)(0,12){3}{\oval(1,6)[r]}
\put(-10,10) {\line(1,0){20}} 
\put(50,10) {\line(1,0){30}} 
\multiput(10,10)(6,0){7} {\line(1,0){4}} 
\put(28,7){$\triangleleft$}
\put(-15,0){$p$}
\put(-15,15){$\alpha$}
\put(70,0){$q$}
\put(70,15){$\beta$}
\put(8,50){$\mu\; k_1$} \put(38,50){$\nu\; k_2$}
\put(2,7){$<$}\put(55,7){$<$} \end{picture}
-\qquad
\begin{picture}(100,60)(0,10)
\multiput(40,15)(-10,10){4}{\oval(10,10)[rt]}
\multiput(50,15)(-10,10){4}{\oval(10,10)[lb]}
\multiput(20,15)(10,10){1}{\oval(10,10)[lt]}
\multiput(10,15)(10,10){2}{\oval(10,10)[rb]}
\multiput(40,45)(10,10){1}{\oval(10,10)[rb]}
\multiput(40,35)(10,10){2}{\oval(10,10)[lt]}
\put(-10,10) {\line(1,0){20}} 
\put(50,10) {\line(1,0){30}} 
\multiput(10,10)(6,0){7} {\line(1,0){4}} 
\put(28,7){$\triangleleft$}
\put(-15,0){$p$}
\put(-15,15){$\alpha$}
\put(70,0){$q$}
\put(70,15){$\beta$}
\put(5,53){$\mu\; k_1$}
\put(38,53){$\nu\; k_2$}
\put(2,7){$<$}\put(55,7){$<$}
\end{picture}
\\
 e^2 M^{(2)}_{\mu\nu\,;\alpha\beta}(k_1,k_2\,;p,q)
&=&  
\begin{picture}(85,50)(0,30)
\multiput(20,34)(-10,10){2}{\oval(10,10)[rt]}
\multiput(30,34)(-10,10){2}{\oval(10,10)[lb]}
\multiput(40,34)(10,10){2}{\oval(10,10)[lt]}
\multiput(30,34)(10,10){2}{\oval(10,10)[rb]}
\multiput(30,13)(0,12){2}{\oval(1,6)[l]}
\multiput(32,19)(0,12){1}{\oval(1,6)[r]}
\put(-10,10) {\line(1,0){90}} 
\put(-15,0){$p$}
\put(-15,15){$\alpha$}
\put(26,15){\footnotesize{$\bigtriangleup$}}
\put(70,0){$q$}
\put(70,15){$\beta$}
\put(5,53){$\mu\; k_1$}
\put(38,53){$\nu\; k_2$}
\put(8,7){$<$}\put(50,7){$<$}
\end{picture}
\qquad 
+\qquad
\begin{picture}(85,50)(0,30)
\multiput(20,34)(-10,10){2}{\oval(10,10)[rt]}
\multiput(30,34)(-10,10){2}{\oval(10,10)[lb]}
\multiput(40,34)(10,10){2}{\oval(10,10)[lt]}
\multiput(30,34)(10,10){2}{\oval(10,10)[rb]}
\multiput(30,13)(0,12){2}{\oval(1,6)[l]}
\multiput(32,19)(0,12){1}{\oval(1,6)[r]}
\put(-10,10) {\line(1,0){90}} 
\put(-15,0){$p$}
\put(-15,15){$\alpha$}
\put(26,18){\footnotesize{$\bigtriangledown$}}
\put(70,0){$q$}
\put(70,15){$\beta$}
\put(5,53){$\mu\; k_1$}
\put(38,53){$\nu\; k_2$}
\put(8,7){$<$}\put(50,7){$<$}
\end{picture}\quad\,.
\\ \\ \\
\eea Explicitly,  
\beaq
 M^{(1)}_{\mu\nu \,;\alpha\beta}(k_1,k_2\,;p,q)
&=& 
\sum_{\delta\,,\gamma} \int_\ell\,
\Big\{ 
v_{\alpha\delta\,;\mu}(p,\ell;k_1) S_{\delta\gamma}^{+R} (\ell)
v_{\gamma\beta\,;\nu}(\ell,q;k_2)
\cr
&&\qquad \times 
(2\pi)^{2D} \delta(p-k_1 -\ell) \delta(\ell-q-k_2) 
+ k_1 \leftrightarrow k_2 \Big\}
\cr
 M^{(2)}_{\mu\nu\,;\alpha\beta}(k_1,k_2\,;p,q)
&=& 
\sum_{\delta\,,\gamma} \int_\ell\,
\Big\{ v_{\alpha\beta\,;\rho}(p,q;\ell) 
v_{\mu\nu\rho}(k_1,k_2,\ell) 
D^R(\ell)
\cr
&&\qquad \times 
(2\pi)^{2D} 
\delta(p+\ell-q) \delta(k_1+k_2+\ell) 
+ \ell \to -\ell  \Big\}\,.
\qquad
\eeaq

The gauge invariance requires the identity 
\beq
k_1^{\mu}\,\epsilon^\nu(\lambda)\, 
\bar u_\alpha(p) \,
 M_{\mu\nu\,;\alpha\beta}(k_1,k_2\,;p,q)\, u_\beta(q)
=0\,.
\eeq
For the commuting case this is trivially satisfied 
since 
$ M^{(2)}_{\mu\nu\,;\alpha\beta}(k_1,k_2\,;p,q) =0 $,
and due to the identity
\bea
\bar u_\alpha(p) \,
k\!\!\!\slash_1 \,
\frac1{\ell\!\!\!\slash -m +i\epsilon} 
\Big|_{p=k_1+\ell}
= -\bar u_\alpha(p)
\\
\frac1{\ell\!\!\!\slash -m +i\epsilon} \,
k\!\!\!\slash_1\, 
u_\beta(q)\Big|_{\ell=k_1+q}
= u_\beta(q)\,,
\eea
we have 
\beq
k_1^{\mu}\,\epsilon^\nu(\lambda)\, 
\bar u_\alpha(p) \,
 M^{(1)}_{\mu\nu\,; \alpha\beta}(k_1,k_2\,;p,q)\, u_\beta(q)
= -\bar u_\alpha(p)\epsilon\!\!\slash (\lambda)\,  
\, u_\beta(q)
+ \bar u_\alpha(p)\epsilon\!\!\slash (\lambda)\,  
\, u_\beta(q)
=0\,.
\eeq

For the non-commuting case, the result is not obvious at all.
First, we note that
\beaq
&& \!\!\!\!\!\! \!\!\!\!\!
M^{(1)}(k_1,k_2\,;p,q)
\equiv k_1^{\mu}\,\epsilon^\nu(k_2, \lambda)\, 
\bar u_\alpha(p) \,
 M^{(1)}_{\mu\nu\,; \alpha\beta} (k_1,k_2\,;p,q)\, u_\beta(q)
\cr
&&  \!\!\!
= e^{\frac i2 (p-q)\wedge \ell }\,
\bar u(p) \,k\!\!\!\slash_1\, 
\Bigg\{
\Bigg( \frac{ \hat \ell\!\!\slash^+ +m}{2 \omega_\ell}\Bigg)\,
\frac i {\ell^0 - \omega_\ell + i \epsilon}\,
-
\Bigg( \frac{\hat \ell\!\!\slash^- +m}{2 \omega_\ell}\Bigg)\,
\frac i {\ell^0 + \omega_\ell - i \epsilon}\,
\Bigg\}
\epsilon\!\! \slash (k_2, \lambda) \, 
u(q)\Bigg|_{\ell= p-k_1}
\cr
&&  \!\!\!\!\!\!
+ 
e^{\frac i2 (p-q)\wedge \ell}\,
\bar u(p)\,
\epsilon\!\! \slash (k_2, \lambda) \, 
\Bigg\{
\Bigg( \frac{\hat \ell\!\!\slash^+ +m}{2 \omega_\ell}\Bigg)\,
\frac i {\ell^0 - \omega_\ell + i \epsilon}\,
-
\Bigg( \frac{\hat \ell\!\!\slash^- +m}{2 \omega_\ell}\Bigg)\,
\frac i {\ell^0 + \omega_\ell - i \epsilon}\,
\Bigg\} \, k\!\!\!\slash_1\, 
u(q)\Bigg|_{\ell=q+k_1} 
\!\!\!\!\!\!\!\!\!\!\!\!\!\!\!
\,.
\label{m_1}
\eeaq
Using the on-shell condition,
\bea
\bar u (p) ( p\!\!\slash -m) &=& ( p\!\!\slash -m) u(p) =0
\cr
(\hat \ell\!\!\slash^+ +m)(\hat \ell\!\!\slash^+  -m)
&=&(\hat \ell\!\!\slash^+ -m)(\hat \ell\!\!\slash^+ +m) 
=0
\cr
(\hat \ell\!\!\slash^- +m)(\hat \ell\!\!\slash^- -m)
&=& (\hat \ell\!\!\slash^- -m)(\hat \ell\!\!\slash^- +m) 
=0
\eea
we may have 
\beaq
\bar u (p) ( \ell\!\!\slash - p\!\!\slash) ( \hat \ell\!\!\slash^+ +m) 
&=&
\bar u (p) \gamma^0 ( \hat \ell\!\!\slash^+ +m)  ( \ell^0 - \omega_\ell) 
\cr
\bar u (p) ( \ell\!\!\slash - p\!\!\slash) ( \hat \ell\!\!\slash^- +m) 
&=&
\bar u (p) \gamma^0 ( \hat \ell\!\!\slash^+ +m)  ( \ell^0 + \omega_\ell) 
\cr
( \hat \ell\!\!\slash^+ +m) ( \ell\!\!\slash - q\!\!\slash) u(q)
&=&
 ( \ell^0 - \omega_\ell) ( \hat \ell\!\!\slash^+ +m)  \gamma^0 u (q) 
\cr
( \hat \ell\!\!\slash^- +m) ( \ell\!\!\slash - q\!\!\slash) u(q)
&=&
 ( \ell^0 + \omega_\ell) ( \hat \ell\!\!\slash^- +m)  \gamma^0 u (q) \,.
\eeaq
Due to this identities, we may put  (\ref{m_1}) as
\beaq
M^{(1)}(k_1,k_2\,;p,q)
= -i \Big( e^{\frac i2 (p-q)\wedge (p-k_1) }\,
-e^{\frac i2 (p-q)\wedge(p-k_2)}\,\Big)\,
\bar u(p)\,
\epsilon\!\! \slash (k_2, \lambda) \, u(q) 
\eeaq

On the other hand, $M^{(2)}$ is given as 
\beaq
&& M^{(2)}(k_1,k_2\,;p,q)
\equiv k_1^{\mu} \, \epsilon^\nu(k_2, \lambda)\, 
\bar u_\alpha(p) \,
 M^{(2)}_{\mu\nu\,; \alpha\beta} (k_1,k_2\,;p,q)\, u_\beta(q)
\cr
&& \qquad
=  e^{\frac i2  p\wedge q } \,
\bar u(p) \, \gamma^\rho \,u(q) \, 
k_1^\mu\,  \epsilon^\nu(k_2, \lambda)\, 
v_{\mu\nu\rho}(k_1, k_2, \ell)\,
 D^R(\ell) \Big|_{\ell=q-p}
\cr
&&  \qquad \qquad
+ e^{\frac i2  p\wedge q } \,
\bar u(p) \, \gamma^\rho \,u(q) \, 
k_1^\mu \, \, \epsilon^\nu(k_2, \lambda)\, 
v_{\mu\nu\rho}(k_1, k_2, -\ell)\,
D^R(\ell) \Big|_{\ell=p-q}\,.
\eeaq
Using the vertex relation,
\bea
k_1^\mu\, v_{\mu\nu\rho}(k_1, k_2, \ell)\,
=-2i
\Bigg(
\sin{\frac{k_1 \wedge \ell}2}
\Big( k_1 \cdot k_2 \,\delta_{\nu \rho}
-k_{2\,\rho}\, k_{1\, \nu} \Big)
+
\sin{\frac{k_2 \wedge k_1}2}
\Big( \ell_\nu \cdot k_{1\, \rho} 
-\ell \cdot k \, \delta_{\nu \rho} \Big)
\Bigg)
\eea
and on-shell conditions of the photon,
\bea
2 k_1 \cdot k_2 = (k_1 + k_2 )^2 \,,\qquad
k_2 \cdot \epsilon (k_2) =0 
\eea
we have 
\beaq
&& \!\!\!\!\!\!\!\!\!\!
M^{(2)}(k_1,k_2\,;p,q)
\cr
&&
=  - e^{\frac i2  p\wedge q } \,
2 \sin{\frac{k_1\wedge k_2}2} 
\Bigg(
\bar u(p) \, \epsilon\!\!/ (k_2, \lambda)) \,u(q)
 \,
+  \frac{ k_1 \cdot \epsilon\!\!/ (k_2, \lambda)}
{(q-p)^2} \,
\bar u(p) ( q\!\!/ -p\!\!/) u(q) 
\Bigg)\,.\quad
\eeaq
Noting that electron on shell satisfies
\bea
\bar u(p) ( q\!\!\!/ -p\!\!\!/) u(q) =0
\eea
we have
\bea
&&M^{(2)}(k_1,k_2\,;p,q)
=  -i \Big( 
e^{\frac i2  (p\wedge q + k_1 \wedge k_2)} \,
-
e^{\frac i2  (p\wedge q - k_1 \wedge k_2)} \Big)
\, \bar u(p) \, \epsilon\!\!\!/ (k_2, \lambda)) \,u(q) \,
\\
&&\qquad\qquad
=-i \Big(
e^{\frac i2  (p-q)\wedge(p-k_1)} \,
-
e^{\frac i2  (p-q)\wedge (p- k_2) } \Big)
\, \bar u(p) \, 
\epsilon\!\!\!/ (k_2, \lambda)) \,u(q) \,.
\eea
Therefore, 
$M^{(1)} $ and $M^{(2)} $ 
cancel exactly each other.
This proves the gauge invariance of the scattering 
amplitude to the lowest order.

We remark that this gauge invariance result is 
in contrast with the result in \cite{nonward}
where the gauge symmetry is seen violated 
to the lowest order  
in the time-ordered perturbation theory 
defined in \cite{topt}.
TOPT differs from ours in the time-ordering:
Our minimal realization of the time-ordering
cures the ill-defined time-ordering and 
preserves the gauge symmetry.

\vskip 0.5cm
\section{Classical solution of the vacuum \label{V}}
\noindent

In this section, we turn to 
the classical vacuum solution of STNC
pure abelian gauge theory.
NC abelian gauge theory is known 
to possess the non-abelian nature and 
self-sustaining magnetic flux 
in Euclidean NCFT \cite{agms}.
One may wonder whether there is a non-trivial finite 
action solution in STNC gauge field theories also. 
Indeed,  it is pointed out in \cite{bk} that 
non-trivial vacuum solutions exist in STNC NC gauge theory
and the validity of STNC perturbative gauge theory 
has been questioned.  We address this problem in connection 
with the gauge fixing condition. 
For simplicity, we consider the 2 dimensional theory
and use the notation, $x^\mu=(t, {\bf x} )$.

The classical equation of NC abelian theory
is given as
\beq
[\,\partial_\mu -ig A_\mu , F^{\mu\,\nu}\,]_\star =0\,. 
\label{cleq}
\eeq
Suppose we choose the temporal gauge $ A_0 =0$.
The field strength simplifies to 
$ F_{01}= \partial_0 A_1 $ 
and satisfies the equation,
\beq
\partial_0 F_{01} =\partial_0^2 A_1 = 0\,.
\eeq
Therefore $A_1$ and $F_{01}$ is of the form,
\beq A_1 (x) = t c ({\bf x}) + d ({\bf x}) \,,\qquad 
F_{01} = c({\bf x} )\,. 
\eeq 
The fields should satisfy the remaining field equation,
\bea
0= \partial_x c({\bf x})  
-ig [t c({\bf x})+ d({\bf x}), c({\bf x})]_\star
=\partial_x c({\bf x}) \Big(1  - e \theta \, c({\bf x})\Big) 
\eea
and the solution is given as 
\beq
F_{01}= c({\bf x})= \,{\rm constant}\, =0 
\eeq
to have the action finite.
The classical solution of the field strength 
with finite action is unique and vanishes.
The same goes with the axial gauge, $ A_1=0$.

On the other hand, 
suppose  one chooses a radial gauge
\bea 
x^\mu A_\mu =0 \,.
\eea 
One may get a pure gauge form satisfying the symmetric gauge
using the gauge transformation.
If we choose the gauge transformation, 
\beq
U(x) = \frac \theta 2
\int \frac{d^2 k}{2\pi}\, e^{-ik \cdot x}
= \pi \theta  \,\,\delta (x) \,,
\label{U}
\eeq
we have the pure gauge form of the field,
\beq
-i g A_\mu = \ U \partial_\mu U^\dagger 
= -\frac 2 \theta \epsilon_{\mu \nu} x^\nu\,.
\eeq
One may obtain a localized pure gauge field
of a symmetric form
\beq
A_\mu = \epsilon_{\mu \nu} x^\nu f( r) 
\eeq
where $f(r)$ is a localized function
with $r^2 = t^2 +x^2\,\,$,
if one applies to $A_\mu=0$,
unitary gauge transformation,
\beq
U_n (x) = e^{i \alpha_n g_{nn} (x)}_\star 
=1 + (e^{i \alpha_n}-1) \,g_{nn}(x)\,.
\eeq
Here $\alpha_n$ is an arbitrary real number 
(conveniently put as $\pi$) and 
$g_{mn}(x)$ is the generalized Wigner function 
(with $x \to \sqrt{\theta} x$ rescaled) \cite{wf}, 
\beq
g_{mn} (x) = 2 e^{-\bar z z } \sqrt{\frac{n!}{m!}} (-1)^n 
(\sqrt2 z)^{m-n} L_n^{m-n} ( 2 r^2) = g_{nm}^*  \qquad {\rm for }\,\,  m\ge n
\eeq
where $L_n^{a} (t)$ is the Laguerre polynomial,
\beq
L_n^{a} (t) = \frac 1{n!} \frac {e^t}{t^a} \frac{d^n}{dt^n} \Bigg(e^{-t} t ^{a+n}\Bigg)
\eeq  
and $z=t+ix$.
Note that $g_{mn}(x)$ satisfies the relation,
\beq
g_{mn} (x) \,\star\, g_{pq} (x) 
= \delta_{n\,p} g_{mq}(x)\,.
\eeq
For example, using 
$ U_0(z)= 1- 2 g_{00}(z)$,
one has the localized form of the pure gauge field,
\beaq
-ig \Big( A_0 -i A_1 \Big) (x) 
&=& U_0^\dagger \star \, \Big( 2\partial \, U_0 \Big)
=  2 \sqrt{\frac 2\theta } \,\, g_{10}(x)
\cr
-ig \Big( A_0 +i A_1 \Big)  (x) 
&=& U_0^\dagger \star \, \Big(2\bar \partial \, U_0 \Big)
=  2 \sqrt{\frac 2\theta} \,\,  g_{10}(x)^* 
\eeaq
where 
$\partial = (\partial_0 -i \partial_1)/2$
and 
$\bar \partial = (\partial_0 +i \partial_1)/2$
and the following identities are used:
\bea
\partial g_{nn}= - \sqrt{\frac{n+1} {2\theta} }\,g_{n\,n+1}
+  \sqrt{\frac{n} {2\theta} }\, g_{n-1\,n}
\,,\qquad
\bar \partial g_{nn}= - \sqrt{\frac{n+1} {2\theta} }\, g_{n+1\,n}
+  \sqrt{\frac{n} {2\theta} }\, g_{n\,n-1}\,.
\eea 
Any composite operator constructed from $U_n$'s 
will give the localized pure gauge form. 

On the other hand, using the so-called shift
operator $K$, which is not unitary,
\bea
K^\dagger\, \star\,K =1 \ne K \,\star\,K^\dagger\,,
\eea
one may construct 
a non-trivial finite action solution. 
For example,
\beq
K_1= \sum_{n=0}^\infty g_{n+1\, n}(x) 
\eeq
is a shift operator with 
$ K_1^\dagger\,\star\, K_1 =1 $ 
and $ K_1 \,\star\,K_1^\dagger = 1 - g_{00}(x)\,\,$.
This results in the gauge field 
\beaq
&&-ig \Big( A_0 -i A_1 \Big) (x) 
= K_1^\dagger \star  \, \Big( 2\partial \, K_1\Big)
=  \sum_{n=1}^\infty 
2 d_n \,\, g_{n\, n-1}(x)
\cr
&&
-ig \Big( A_0 +i A_1 \Big) (x) 
= K_1^\dagger \star  \, \Big( 2\bar \partial \, K_1\Big)
=  \sum_{n=1}^\infty 
2 d_n \,\, g_{n-1\, n}(x)\,.
\label{localgauge}
\eeaq
Here $ d_n=\sqrt{\frac n {2\theta} } 
- \sqrt{\frac{n+1}  {2\theta} }\,\,$
and the  following identities are used:
\bea
\partial g_{n+1\,n}=  \sqrt{\frac{n+1} {2\theta} }\,
\Big( g_{nn}-g_{n-1\,n+1}
\Big)
\,,\qquad
\bar \partial g_{n+1\,n}= - \sqrt{\frac{n+2} {2\theta} }g_{n+2\,n}
+  \sqrt{\frac{n} {2\theta} } \, g_{n+1\,n-1}\,.
\eea
The gauge function (\ref{localgauge})
is localized and convergent.
This gauge field gives the localized field strength,
$ F_{01}(x) = -g_{00}(x)/(g \theta) $
whose action is finite. 
If one uses a composite operation,
$ V(x) \equiv U (x) \star K_1 (x) $ with $U(x)$ in (\ref{U}),
one then obtains 
\bea
&&-ig \Big( A_0 -i A_1 \Big) (x) 
= V^\dagger \star  \, \Big( 2\partial \, V\Big)
=  -\bar z - \sum_{n=0}^\infty 
\sqrt{\frac{2 n}\theta } \,\, g_{n\, n+1}(x)
\cr
&&-ig \Big( A_0 +i A_1 \Big) (x) 
= V^\dagger \star  \, \Big( 2\bar \partial \, V\Big)
=   - z - \sum_{n=0}^\infty 
\sqrt{\frac{2 n} \theta} \,\, g_{n+1\, n}(x) \,.
\eea

Finally, one may consider the Lorentz gauge 
\bea
\partial^\mu A_\mu =0 \,.
\eea
In this gauge, however, 
there is no such localized solution to have
the finite action. 
The above analysis suggests that the
choice of the gauge fixing may allow 
a different class of solutions. 
One may look for a singular gauge transformation 
which transforms the localized finite action solution
in a radial gauge into a pure gauged solution 
in a axial gauge.
So far, we could not succeed in finding this 
transformation.

The gauge fixing dependence of the solution 
is not limited to STNC but 
the same problem also arises in the SSNC case
(Euclidean case).
Thus, it remains to be seen how the 
gauge fixing condition affects the Hilbert space 
on the NCFT in general. 

\vskip 1cm 

\section{Conclusion \label{L}}

We construct the unitary S-matrix for the space-time
non-commutative abelian gauge theory and QED.
The Feynman rule is presented, based on this S-matrix.

The gauge invariance of the STNC QED perturbation 
is explicitly checked to the lowest order 
using the Compton scattering process.
In this paper we choose the Feynman gauge $\lambda=1$ 
to avoid the unnecessary complication. For other choice
of gauge, one may accommodate the ghost 
to make the S-matrix gauge invariant.  
To prove the gauge symmetry to all orders of 
perturbation theory, one may refer to the 
BRST symmetry as done in SSNC QED \cite{brst}.

Finally, the classical vacuum solution demonstrates 
that the Hilbert space may differ depending on the 
choice of gauge. As seen in the text, the axial gauge 
condition and the radial gauge condition give the 
different class of vacuum solutions. The 
relation between the two different Hilbert spaces
is not clearly understood yet. 
 
\vskip 1cm

{\sl Acknowledgement:}  
It is acknowledged that this work was supported in part 
by the Basic Research Program of the Korea Science 
and Engineering Foundation Grant number 
R01-1999-000-00018-0(2003)(CR), and by Korea Research Foundation 
under project number KRF-2003-005-C00010 (JHY).
CR is also grateful for KIAS while this work is made during his visit.

\end{document}